\documentclass[prl,showpacs,twocolumn]{revtex4}
\usepackage{graphicx}
\usepackage{amssymb}
\usepackage{color}
\usepackage{rotating}

\newcommand{\dd}{\mbox{d}}
\newcommand{\nn}{\nonumber} 
\newcommand{\Eqref}[1]{Eq.~(\ref{#1})}
\newcommand{\ket}[1]{\left|{#1}\right\rangle}
\newcommand{\bra}[1]{\left\langle{#1}\right|}
\newcommand{\scalar}[2]{\left.\left\langle{#1}\right|{#2}\right\rangle}
\newcommand{\tfrac}[2]{\mbox{$\frac{{#1}}{{#2}}$}}

\begin{document}

\title{Collisional Quantum Brownian Motion}

\author{I. Kamleitner} \author{J. Cresser} 
\affiliation{Centre for Quantum Computer Technology, Physics Department, Macquarie University, Sydney, New South Wales 2109, Australia}

\begin{abstract} 
	We derive a quantum master equation from first principles to describe friction in one dimensional, collisional Brownian motion. We are the first to avoid an ill-defined square of the Dirac delta function by using localized wave packets rather than plane waves. Solving the Schr\"odinger equation for two colliding particles, we discover that the previously found position diffusion is not a physical process, but an artifact of the approximation of a coarse grained time scale.
\end{abstract}
\pacs{05.20.Dd,03.65.Yz,03.75.-b,47.45.Ab}
\maketitle

More than a hundred years ago, Einstein and Smoluchowski \cite{Einstein} derived a successful classical description of  Brownian motion, the erratic movement of a particle affected by ambient gas molecules. In the last few decades, the field of quantum Brownian motion (QBM) has become established~\cite{Joos},~\cite{Caldeira}. Not only is such a quantum description of friction and diffusion desirable, but QBM also sheds light on the non-classical process of decoherence which is believed to be of importance in the quantum classical transition.

The field of QBM is still the source of open questions. Of particular concern is deriving a valid master equation (ME) for the reduced density operator of the Brownian particle in the Markov limit, in which case a ME of Lindblad form is to be expected. What  should be noted here is that all such MEs include position diffusion. This process is impossible in classical dynamics (although the position does diffuse in the long term due to randomness of the momentum) as it derives from position jumps which are at odds with continuous trajectories of classical particles. Therefore it is currently believed that position diffusion arises as a quantum effect accompanying collisional quantum friction \cite{review}. In this letter we show the contrary: We find that position diffusion is not a real physical process in collisional QBM, but results from using a coarse grained time scale on which collisions appear instantaneous. Unproblematic in classical dynamics where collision times of hard-core particles are indeed short, the assumption of instantaneous collisions has to be used with care in quantum dynamics, where collision times depend on the widths of the colliding wave packets.

There are several ``non-collisional" models of QBM in the literature (see e.g. \cite{Hu}-\cite{Jacobs}). Here we concentrate on collisional models which are the closest in spirit to the original Einstein approach, and which can be tested in experimental setups~\cite{Exp}. Important contributions in the literature include \cite{Diosi}-\cite{Hornberger2} and can be reviewed in~\cite{review}. Mostly, the analysis is based on using scattering theory to describe the effects of a single collision.

To point out some problems which usually occur in collisional QBM when using scattering theory, we briefly review Hornberger's contribution~\cite{Hornberger2} as it seems to be the most complete one. 
Due to momentum conservation a two particle collision reduces to a one particle problem, where the transition operator $T$ with momentum matrix elements related to the scattering amplitude $\langle\mathbf p'|T|\mathbf p\rangle = \delta(\mathbf p' \phantom{\!\!l}^{\!2}-\mathbf p^2)f(\mathbf p',\mathbf p)/(\pi\hbar)$ is of importance. In particular it appears in an expression of the form 
	\begin{eqnarray}
		X=\frac{(2\pi\hbar)^3}\Omega \langle\mathbf p' + \mathbf q |T|\mathbf p+\mathbf q\rangle  \langle\mathbf p'-\mathbf q|T^\dag|\mathbf p-\mathbf q\rangle, \quad \label{problem}
	\end{eqnarray}
where $\Omega\to\infty$ is a box-normalization volume. For $\mathbf q=0$ this term contains an ill-defined square of the Dirac delta function and a physically motivated replacement rule
	\begin{eqnarray}
		\frac{(2\pi\hbar)^3}{\Omega}|\langle\mathbf p'|T|\mathbf p\rangle |^2 &\to& \delta\!\left(\frac{\mathbf p' \phantom{\!\!l}^{\!2}-\mathbf p^2}2\right) \frac{|f(\mathbf p',\mathbf p)|^2}{\sigma(\mathbf p)|\mathbf p|}.\qquad \label{replacement}
	\end{eqnarray}
is used. However, for $\mathbf q \neq 0$ \Eqref{problem} is well defined. In fact, using standard relations for Dirac delta functions,
	\begin{eqnarray}
		X &=& \frac{\pi\hbar}{\Omega p q} \delta(p-p')\delta\left((\mathbf p-\mathbf p')_{\|\mathbf q}\right) \nn\\
		&&  \times \; f(\mathbf p'+\mathbf q,\mathbf p+\mathbf q) f^*(\mathbf p'-\mathbf q,\mathbf p-\mathbf q)\qquad \label{correct}
	\end{eqnarray}
is found, where $q=|\mathbf q|$ and $\mathbf p_{\|\mathbf q}$ is $\mathbf p$ projected on $\mathbf q$. Nevertheless, this result  \Eqref{correct} is not without its problems as the off-diagonal matrix elements $\langle \mathbf p '|\rho |\mathbf p \rangle$ of the Brownian particle's density matrix would be found to disappear in a single collision due to $\Omega\to\infty$, no matter how small $\mathbf p -\mathbf p '$. This seemingly strange observation comes with the fact that the collision time with a gas particle in a non-localized momentum eigenstate is infinite.

To construct a Markovian (memory less) ME, Hornberger assumed instantaneous collisions. But within this assumption, the complete vanishing of coherences is not acceptable any more. To get around this problem, Hornberger substituted the square root of the replacement rule \Eqref{replacement} for $\langle\mathbf p'\pm \mathbf{q}|T|\mathbf p\pm\mathbf{q}\rangle$ in \Eqref{problem} to effectively bring back the otherwise lost coherences of momentum states.

The replacement rule~(\ref{replacement}) has to be used with caution in the first place, but its application to a well defined expression is even more questionable. Whether decoherence in momentum bases and the closely related position diffusion are correctly described by approaches of this kind is surely not certain.

To resolve this matter we study a single collision in terms of localized gas states. Then a collision time can be precisely defined, and the low-density and high-temperature (LDHT) limit will be quantified by `\emph{collision time}' $\times$ `\emph{collision rate}' $\ll 1$. To examine a collision event time resolved, and to investigate under which conditions a complete collision of two wave packets occurs, we analytically solve the two-particle Schr\"odinger equation. This is in contrast to scattering calculations which postulate a complete collision. 

To simplify the problem, we consider one-dimensional QBM and assume a hard core interaction potential between gas and Brownian particles. Our notations are as follows.  Position and momentum operators are denoted with a hat, whereas wave function variables are primed. The index $g$ refers to gas particles.

The two particle Hamiltonian
\begin{equation}
	H=\frac{\hat{p}_g^2}{2m_g}+\frac{\hat{p} ^2}{2m }+a\delta(\hat{x} -\hat{x}_g),\quad a\to\infty\label{ham}
\end{equation}
forbids any tunneling of the gas particle through the Brownian one. The two orthogonal sets of energy eigenstates in position space are
	\begin{eqnarray}
		\psi^a_{\tilde k\bar k}(x_g',x ') &=& e^{i\bar k(x '+\alpha x_g')}\sin(\tilde k(x_g'-x ')),\quad \label{3}\\
		\psi^s_{\tilde k\bar k}(x_g',x ') &=& e^{i\bar k(x '+\alpha x_g')}\sin(\tilde k|x_g'-x '|),\label{4}
	\end{eqnarray}
with $\bar k\in \mathbb{R},\;\tilde k\in\mathbb{R}_+$, $\alpha=m_g/m $, and energy $E(\tilde k,\bar k)=(1+\alpha)(\tilde k^2+\alpha\bar k^2)\hbar^2/(2m_g)$.

The collisional analysis is based on noting that the density operator of a thermal ideal gas particle at temperature $T$ can be written as convex decomposition of \emph{Gaussian} states
	\begin{equation}
		\rho_g = \int \frac{\dd x_g}{L} \int \dd p_g\, \mu_{\sigma_g}(p_g) \ket{x_g,p_g}_{\!\sigma_g}\!\!\!\bra{x_g,p_g}, \label{thermalgas}
	\end{equation}
where $L$ is a normalization length and where $\ket{x,p}_{\!\sigma_g}$ denotes a Gaussian wave packet with position variance $\sigma_g$: 
		\begin{equation}
			\scalar{x'}{x,p}_{\!\sigma_g}= \frac{e^{-ixp/2\hbar}}{\sqrt{\sqrt\pi \sigma_g}} e^{ix'p/\hbar}e^{-(x-x')^2/2\sigma_g^2} \label{gauss}
		\end{equation}
Hornberger and Sipe~\cite{Hornberger} showed that \Eqref{thermalgas} is valid if $\mu_{\sigma_g}(p_g) = \frac{1}{\sqrt{2\pi m_g k_BT_{\sigma_g}}} e^{-p_g^2/2m_gk_BT_{\sigma_g}}$ and $T_{\sigma_g}=T-\frac{\hbar^2}{2m_gk_B\sigma_g^2}$.

We therefore undertake this analysis by considering an initial state that approaches the product state
	\begin{equation}
		\ket{x_g,p_g}\otimes\ket{x ,p } \label{init}
	\end{equation}
in the limit of large separation, i.e. we assume vanishing overlap of the two initial wave packets
	\begin{eqnarray}
		|x_g-x |\gg \sqrt{\sigma^2_g+\sigma^2 }.\label{overlap}
	\end{eqnarray}

Thus we have to construct an initial two-particle state, expanded in terms of energy eigenstates \Eqref{3} and (\ref{4}), which satisfies the boundary condition $\psi(t=0, x_g, x =x_g)=0$ required by the hard core interaction, and which approaches \Eqref{init} in the limit \Eqref{overlap}. As the procedure is quite lengthy, we only present the result and refer to our more detailed paper \cite{PRA}. It may be checked that
	\begin{eqnarray}
		\psi(0,x_g',x ') &=& \frac{i}{\sqrt{2}\pi}\int_\infty^\infty \!\! \dd\bar k\! \int_0^\infty \!\! \dd \tilde k\,  \widetilde{\psi}(0,\tilde k,\bar k) \nn\\
		&&\times \left(\psi_{\tilde k\bar k}^a(x_g',x ')-\psi_{\tilde k\bar k}^s(x_g',x ') \right) \qquad\label{10}
	\end{eqnarray}
with
	\begin{eqnarray}
		\widetilde{\psi}(0,\tilde k,\bar k) &=&  \frac{(1+\alpha)\sigma }{\sqrt{2\pi\sqrt{\alpha}}}  \exp\!\left[i\frac{x p (1+\alpha)}{2\alpha\hbar} -\frac{\bar k^2}{2}(1+\alpha)\sigma ^2 \right] \nn\\
		&\times& \!\left\{e^{i\tilde kx (1+\alpha)/\alpha} \exp\!\left[ -\left(\tilde k+\frac{p }{\hbar}\right)^2 \frac{1+\alpha}{2\alpha} \sigma ^2 \right] \right. \nn\\
		&& \!\left. - e^{-i\tilde kx (1+\alpha)/\alpha} \exp\!\left[ -\left(\tilde k-\frac{p }{\hbar}\right)^2 \frac{1+\alpha}{2\alpha} \sigma ^2 \right]\right\}  \nn
	\end{eqnarray}
has indeed the required properties. Without loss of generality we used the reference frame of centre of mass (i.e. $p =-p_g$ and $x =-\alpha x_g$) as well as $x_g<x $ and $p_g>p $. Furthermore, we related the widths $\alpha \sigma_g^2=\sigma ^2$ of the Gaussian wave packets to the relative mass of the colliding particles. This turns out to be the crucial requirement for the two particles not being entangled after the collision (see also \cite{Schmuser}), and $\ket{x,p}_\sigma$ can be seen as the pointer bases of a measurement performed by a gas particle in the state $\ket{x_g,p_g}_\sigma$ \cite{PRA}. 

The time evolution is now easily achieved by multiplying the integrand of \Eqref{10} with $e^{-iE(\tilde k,\bar k)t/\hbar}$. After straight forward integration we find
	\begin{eqnarray}
		\psi(t,x_g',x ') &=&  \frac{\sigma \sqrt{\sqrt{\alpha}}}{\sqrt{\pi}\left(\sigma ^2+\frac{i\hbar t}{m }\right)}     \exp\!\left[ i\frac{x_gp_g+x p }{2\hbar}\right] \nn\\
		\times\hspace{10mm}&&  \hspace{-15mm} \exp\!\left[   - \frac{ \left(\frac{1+\alpha}{\alpha}\right)\sigma ^2\left(x +\frac{p t}{m }\right)^2  +  \left(x'^2+\alpha x'^2_g\right)\!\left(\sigma ^2-\frac{i\hbar t}{m }\right)}    {2\left(\sigma ^4+\frac{\hbar^2t^2}{m ^2}\right)}    \right]\!\! \nn\\
		\times\hspace{10mm}&&  \hspace{-15mm} \left\{  \exp\!\left[  -(x_g'-x ')\frac{\sigma ^2\left(x +\frac{p t}{m }\right)+i\left(\frac{p \sigma ^4}{\hbar}-\frac{x \hbar t}{m }\right)}    {\left(\sigma ^4+\frac{\hbar^2t^2}{m ^2}\right)}  \right] \right.  \nn\\  
		\hspace{10mm}&&  \hspace{-15mm} - \left.  \exp\!\left[ (x_g'-x ') \frac{\sigma ^2\left(x +\frac{p t}{m }\right)+i\left(\frac{p \sigma ^4}{\hbar}-\frac{x \hbar t}{m }\right)}    {\left(\sigma ^4+\frac{\hbar^2t^2}{m ^2}\right)}  \right]   \right\} \label{12}
	\end{eqnarray}
for $x_g'<x'$ and $\psi(t,x_g',x ')=0$ otherwise. At time $t=0$ the first term in the curly brackets is the dominant one. If relation (\ref{overlap}) holds, then we can neglect the second term and \Eqref{12} becomes the position representation of the desired initial state (\ref{init}). If the particles' relative velocity is large compared to their velocity uncertainty
	\begin{equation}
		|p_g|\gg \sqrt{\frac1{1+\alpha}}\frac\hbar{\sigma_g},\label{momoverlap}
	\end{equation}
then we can neglect the first term sufficiently long after the collision. In this case \Eqref{12} is the position representation of the remarkable simple product state 
	\begin{eqnarray}
		U_g(t)\ket{- x_g,- p_g}\otimes U (t)\ket{- x ,- p },
	\end{eqnarray}
where $U(t)$ is the free particle evolution operator. Using \Eqref{12} we can now define the collision time
	\begin{equation}
		t_c=\sqrt{\frac{8}{1+\alpha}}\sigma_g\frac{m_g}{|p_g|}\label{collisiontime}
	\end{equation}
as the time it takes from the first term in the curly brackets to dominate, to the second term to dominate. 

\begin{figure}[htbp]
	\includegraphics[width=\linewidth]{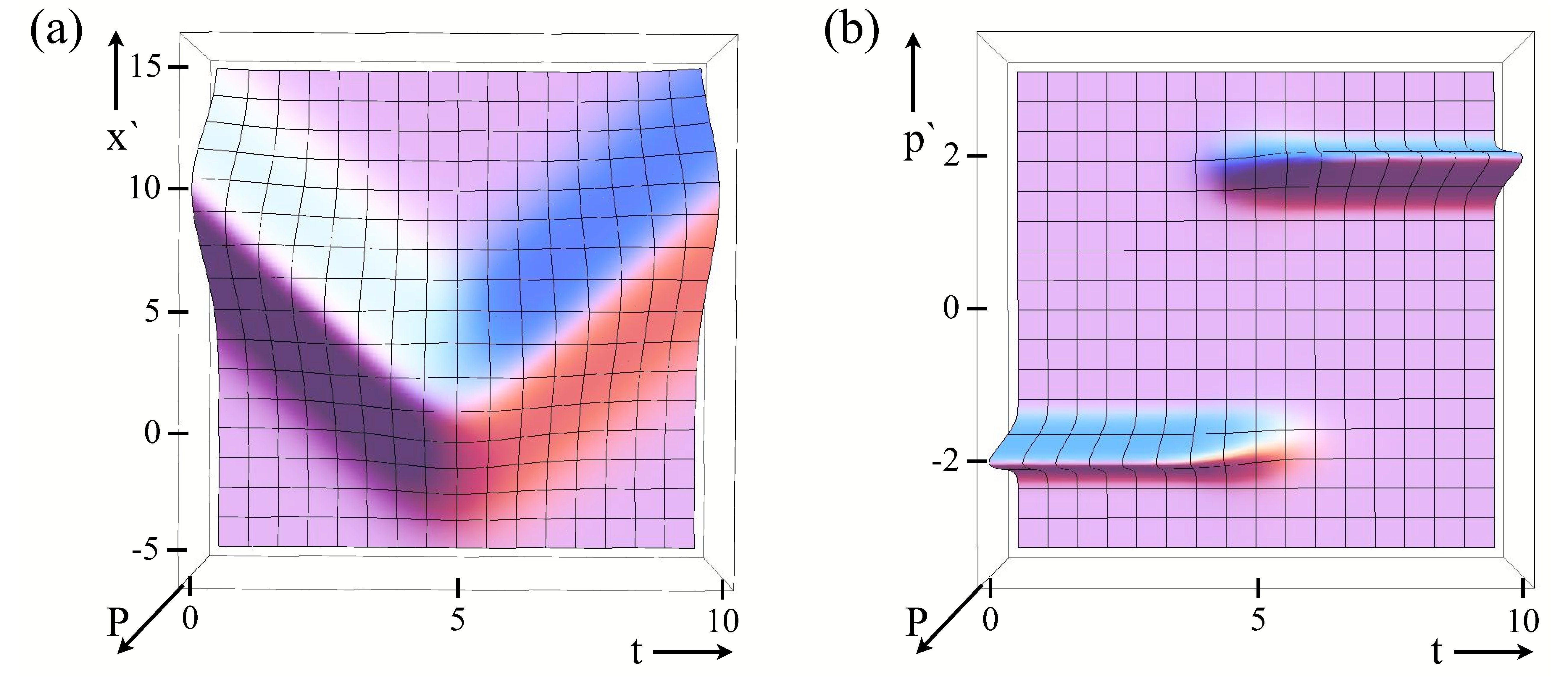}
\caption{\small (color online) Position (a) and momentum (b) probability distributions for the Brownian particle during a collision. While the position probability distribution ``flows", the momentum distribution ``jumps" from the initial value to the final one. This disproves position diffusion in collisional QBM.   Parameters are: $x = 10$, $p = -2$, $m = 1$, $\alpha = 0.3$, $\hbar = 1$, and $\sigma = 4$.}
\label{fig1}
\end{figure}

To examine the behavior of the Brownian particle during a collision in detail we trace out the gas particle from \Eqref{12}. Its position probability distribution $p (t,x') = \int\dd x_g'\,\left|\psi(t,x_g',x ')\right|^2$ can be obtained analytically in terms of the error function and does not show any position jumps. An example is plotted in Fig.~\ref{fig1}(a), and reminds of a classical collision. As Gaussian states build an over complete set of states, this result shows that no position jumps, and hence no position diffusion can occur in collisional QBM.
Using the Fourier transformation of \Eqref{12} with respect to $x'$, we find the momentum distribution of the Brownian particle which shows the expected momentum jump (see Fig.~\ref{fig1}(b)).

In a general reference frame the collision event with Gaussian states can be written as
	\begin{equation}
		\ket{x_g,p_g}\otimes\ket{x ,p } \;\parbox{1.5cm}{$\,${\scriptsize scattering}\\$\overrightarrow{\phantom{scatterin}}$} \; U_g(t)\!\ket{\bar{x}_g,\bar{p}_g}\otimes U (t)\! \ket{\bar{x} ,\bar{p} }  \label{one}
	\end{equation}
where positions and momenta after the collision relate with the initial values in the classical way~\cite{footnote}: 
\begin{eqnarray}
		&&\hspace{-9mm}\bar{x}_g = \frac{2x -(1-\alpha)x_g}{1+\alpha}  ,\hspace{5.8mm} \bar{x}  = \frac{2\alpha x_g+(1-\alpha)x }{1+\alpha}, \quad \label{two}\\
		&&\hspace{-9mm}\bar{p}_g = \frac{2\alpha p -(1-\alpha)p_g}{1+\alpha} , \hspace{4.5mm} \bar{p}  = \frac{2p_g+(1-\alpha)p }{1+\alpha}.\label{three}
	\end{eqnarray}
The transformation of a general Brownian particle's density matrix due to a collision with a gas particle in the state $\ket{x_g,p_g}$ can be written in terms of Kraus operators
	\begin{equation}
		\rho(t) = \int\!\!\!\int\dd \tilde x\dd\tilde p\, U(t)K_{x_g,p_g}(\tilde x,\tilde p)\rho(0)K_{x_g,p_g}^\dagger(\tilde x,\tilde p)U^\dag(t) . \label{Trafo}
	\end{equation}
In \cite{PRA} we will present a derivation of the Kraus operators
	\begin{eqnarray}
		K_{x_gp_g}(\tilde{x},\tilde{p}) \,=\,  \widehat{D}\!\left( \frac{2\alpha}{1\!+\!\alpha}(x_g\!-\!\tilde{x}), \frac{2}{1\!+\!\alpha}(p_g\!-\!\alpha\tilde{p}) \!\right)\!\sqrt{\hat{\pi}(\tilde{x},\tilde{p})}   \nn
	\end{eqnarray}
where $\widehat D(x,p)=e^{i(p\hat x-x\hat p)/\hbar}$ is the Glauber displacement operator and 
	\begin{eqnarray}
		\hat\pi(\tilde x,\tilde p) &=&  \int\!\!\!\int \frac{\dd x\dd p}{2\pi\hbar}w(x,p)\ket{\tilde x+x,\tilde p+p}\!\bra{\tilde x+x,\tilde p+p} \nn\\
		w(x,p) &=&  \frac{2\alpha}{\pi\hbar(1-\alpha)^2} \exp\!\left[ -\frac{2\alpha}{(1-\alpha)^2} \!\left(\frac{x^2}{ \sigma^2}+\frac{ \sigma^2p^2}{\hbar^2}\right) \right]\! \nn
	\end{eqnarray}
are effect operators corresponding to a measurement which is indirectly performed by the gas particle~\cite{PRA}.  The transformation (\ref{Trafo}) may be confirmed by application to Gaussian states and comparison to \Eqref{one}.

Next we determine whether during a time interval $\delta$ the Brownian particle actually collides with a gas particle which is in the state $\ket{x_g,p_g}$. For this purpose we assume 
	\begin{equation}
		\delta\gg t_c \label{coarse}
	\end{equation}
so that we can neglect partial or uncompleted collisions. Then we find that the two particles collide if $(x ,p )$ is within the phase space region $S(x_g,p_g)=\big\{(x ,p )\big|0<\frac{x-x_g}{p_g/m_g-p/m}<\delta \big\}$. Therefore we have to \emph{project} the Brownian particle's density operator onto $S(x_g,p_g)$ before applying the transformation \Eqref{Trafo}. Of course, phase space projections do not exist in quantum mechanics, but as long as (\ref{coarse}) is satisfied we can use the approximate projection operator
	\begin{equation}
		\Gamma_{\delta}(x_g,p_g) =  \int\!\!\!\int_{S(x_g,p_g)} \frac{\dd x\dd p}{2\pi\hbar} \ket{x,p}\!\bra{x,p}. \label{Project}
	\end{equation}

By multiplying this operator with the probability of finding a gas particle in the state $\ket{x_g,p_g}$, we define the collision probability operator
	\begin{equation}
		P_{\delta}(x_g,p_g) = n_g\mu(p_g)  \Gamma_{\delta}(x_g,p_g), \label{Rate}
	\end{equation}
where $n_g$ is the gas number density. In fact, the expectation value Tr$[\rho P_{\delta}(x_g,p_g)]$ gives the probability of a collision of the Brownian particle and a gas particle described by $\ket{x_g,p_g}$. We also define $P_\delta=\int\!\!\int \dd x_g\dd p_g P_{\delta}(x_g,p_g)$ corresponding to the total collision probability.

To find the density operator for the Brownian particle at time $\delta$, quantum trajectory theory \cite{Gardiner} suggests to multiply each Kraus operator by the square root of the probability operator. We use the low density assumption
	\begin{equation}
		\mbox{Tr}(\rho P_\delta)\ll 1,\label{density}
	\end{equation}
to neglect two or more collisions during $\delta$ and find
	\begin{eqnarray}
		\rho(\delta) &=& U(\delta) \left\{ \int\!\!\!\!\int\!\!\!\!\int\!\!\!\!\int \dd x_g\dd p_g\dd \tilde x\dd\tilde p \Big[ K_{x_g,p_g}(\tilde x,\tilde p)\Big. \right.\nn\\
		&& \times \sqrt{P_{\delta}(x_g,p_g)}\rho(0)\sqrt{P_{\delta}(x_g,p_g)}K_{x_g,p_g}^\dagger(\tilde x,\tilde p)\Big]\! \nn\\
		&&  +\bigg. \;\sqrt{1-P_\delta}\rho(0)\sqrt{1-P_\delta} \bigg\} \,U^\dag(\delta). \label{discrete}
	\end{eqnarray}
	
The usual approach to a ME is to expand all operators in $\delta$ and to take the limit $\delta\to 0$. However, this limit is not allowed in $\Gamma_\delta(x_g,p_g)$ as then uncompleted collisions were not negligible. A finite $\delta$ will ultimately lead to errors which we minimize by performing the transition to continuity in the interaction picture and to second order in $\delta$
	\begin{equation}
		\frac{\rho_{int}(\delta)-\rho_{int}(0)}{\delta} \;\approx \;\frac{\rho_{int}\left(\frac{\delta}2+\dd t\right)-\rho_{int}\left(\frac{\delta}2\right)\! }{\dd t} . \label{continuity}
	\end{equation}
Transforming back to the Schr\"odinger picture we find
	\begin{eqnarray}
		\frac{\dd \rho(t)}{\dd t} &=& -\frac i\hbar \left[ H,\rho(t)\right] -\frac12 [R\rho(t)+\rho(t)R] \nn\\
		&&+\; \int\!\!\!\! \int\!\!\!\! \int\!\!\!\! \int\dd x_g\dd p_g\dd\tilde x\dd\tilde p\Big[ U\!\left(\tfrac{\delta}2\right)\! K_{x_g,p_g}(\tilde x,\tilde p) \quad\Big.\nn\\
		&& \times \; \sqrt{ R_{\delta}(x_g,p_g)}U^\dagger\!\left(\tfrac{\delta}2\right)\! \rho(t)U\!\left(\tfrac{\delta}2\right)\! \sqrt{R_{\delta}(x_g,p_g)}\nn\\
		&&\times\; \Big. K^\dagger_{x_g,p_g}(\tilde x,\tilde p) U^\dagger\!\left(\tfrac{\delta}2\right)\Big]  \label{ME}
	\end{eqnarray}
where $R=P_\delta/\delta$ and $R_{\delta}(x_g,p_g)=P_{\delta}(x_g,p_g)/\delta$ are rate operators corresponding to collision rates. This master equation of Lindblad form is our main result. It describes collisional friction as well as decoherence for general mass ratios and velocities, and allows a clear interpretation in terms of quantum trajectories. 

An interesting feature of \Eqref{ME} is the dependence on $\delta$ and $\sigma_g$. These parameters persist in the ME because they are bound from above and below by the relations (\ref{overlap}) and (\ref{momoverlap}), and (\ref{coarse}) and (\ref{density}), respectively and therefore the limits $\delta\to 0$ and $\sigma_g\to \infty$ are not allowed. However, we show in \cite{PRA} that the effects of $\delta$ and $\sigma_g$ on the dynamics vanish in the LDHT limit
	\begin{eqnarray}
		\sqrt{2}(1+\alpha)\frac{n_{gas}\hbar}{\sqrt{\pi m_gk_BT}} &\ll&1,
	\end{eqnarray}
which is also the crucial condition to simultaneously satisfy relations (\ref{overlap}), (\ref{momoverlap}), (\ref{coarse}), and (\ref{density}) for most thermal gas momenta. Our ME is valid on a coarse grained time scale $\delta\gg(1+\alpha)\hbar/(k_BT)$.

We apply our ME to derive equations of motion for expectation values of position and momentum. In the limit of a slow Brownian particle $|p/m|\ll \sqrt{k_BT/m_g}$ (the generalization to fast Brownian particles is found in \cite{PRA} and includes non-linear friction) we find 
	\begin{eqnarray}
		\frac{\dd \langle \hat x \rangle}{\dd t} &=& \frac{\langle\hat p\rangle}{m}, \label{x} \\
		\frac{\dd\langle\hat p\rangle}{\dd t} &=& -f \langle\hat p\rangle, \label{p} \\
		\frac{\dd \left\langle\hat x^2\right\rangle}{\dd t} &=& \frac{\langle \{\hat x,\hat p\} \rangle}{m} + \frac{n_{gas}}{3\sqrt\pi}\left(\frac{2k_BT}{m_g}\right)^{3/2}  \delta^2, \qquad \label{xx}  \\
		\frac{\dd \langle\{\hat x,\hat p\}\rangle}{\dd t} &=& \frac{2\left\langle \hat p^2\right\rangle}{m} - f \langle\{\hat x,\hat p\}\rangle . \label{xp} \\
		\frac{\dd\left\langle\hat p^2\right\rangle}{\dd t} &=& 2f \!\left[  mk_BT-\left\langle \hat p^2\right\rangle \right] , \label{pp}
	\end{eqnarray}
where $f=4n_{g}\sqrt{2m_gk_BT}/(\sqrt\pi m)$ is the friction constant. The second term in \Eqref{xx} represents position diffusion. However, as shown in our single collision calculation this is not a true physical process. It rather results from neglecting partial or uncompleted collisions, which is necessary for any Markovian description of collisional QBM. This is confirmed by the observation that this term is small in the range of validity of our ME.

Using only first principles we derived a quantum ME of Lindblad form for collisional QBM in one dimension in the LDHT limit. We showed that the counter intuitive position diffusion found in earlier work results from approximations necessary to derive Markovian dynamics, but is not a real physical process. The application to expectation values, where we recovered the expected classical equations served as a test of our ME. In a future paper we will use our ME to study the important issue of collisional decoherence which is now in reach of quantitative experiments~\cite{Exp}.


\begin{thebibliography}{99}

\bibitem{Einstein} A. Einstein, Ann. d. Physik \textbf{17}, 549 (1905); M. von Smoluchowski, Ann. d. Physik \textbf{21}, 756 (1906).
\bibitem{Joos} E. Joos and H. D. Zeh, Z. Phys. B: Condens. Matt. \textbf{59}, 223 (1985).
\bibitem{Caldeira} A. O. Caldeira and A. J. Leggett, Physica A \textbf{121}, 587 (1983).
\bibitem{review} B. Vacchini and K. Hornberger, to appear in Phys. Rep. (2009). (doi:10.1016/j.physrep.2009.06.001)
\bibitem{Hu}  B. L. Hu, J. P. Paz, and Y. Zhang, Phys. Rev. D \textbf{45}, 2843 (1992).
\bibitem{Gallis2} M. R. Gallis, Phys. Rev. A \textbf{48}, 1028 (1993).
\bibitem{Jacobs} K. Jacobs, EPL \textbf{85}, 40002 (2009).
\bibitem{Exp} K. Hornberger et. al., Phys. Rev. Lett. \textbf{90}, 160401 (2003).
\bibitem{Diosi} L. Di\'osi, Europhys. Lett. \textbf{30}, 63 (1995).
\bibitem{main} S. M. Barnett and J. D. Cresser, Phys. Rev. A \textbf{72}, 022107 (2005).
\bibitem{Vacchini} B. Vacchini, Phys. Rev. Lett. \textbf{84}, 1374 (2000); B. Vacchini, Phys. Rev. E \textbf{63}, 066115 (2001).
\bibitem{Hornberger2}  K. Hornberger, Phys. Rev. Lett. \textbf{97}, 060601 (2006).
\bibitem{Hornberger} K. Hornberger and J. E. Sipe, Phys. Rev. A \textbf{68}, 012105 (2003).
\bibitem{PRA} I. Kamleitner and J. D. Cresser, to be published.
\bibitem{Schmuser} F. Schm\"user and D. Janzing, Phys Rev A \textbf{73}, 052313 (2006).
\bibitem{Gardiner} C. W. Gardiner, A. S. Parkins, and P. Zoller, Phys. Rev. A \textbf{46}, 4363 (1992).
\bibitem{footnote} Noting that $U(t)$ shifts $x$ to $x+pt/m$.


\end{thebibliography}
\end{document}